\documentclass{basi}
\usepackage{epsfig}
\usepackage{txfonts}
\begin{document}
\title[Diffuse Interstellar Clouds]{Kinematics of Diffuse Interstellar Clouds : \\ 
                         Recent GMRT Results} 
\author[Dwarakanath]%
       {K. S. Dwarakanath\thanks{e-mail:dwaraka@rri.res.in} \\ 
        Raman Research Institute, Bangalore 560 080}
\maketitle
\label{firstpage}

\begin{abstract}
A high latitude HI 21 cm-line absorption survey toward extragalactic sources
was recently completed using the Giant Meterwave Radio Telescope (GMRT). 
A total of 104 sources with $\mid$ b $\mid >$ 15 $^{o}$ and with a 21 cm flux density
greater than 1 Jy were observed for $\sim$ 120 hours. With an optical depth
detection limit of $\sim$ 0.01 this is the most sensitive high-latitude
survey as yet. Most of the detected HI 21 cm-line absorption features
belong to a population with a velocity dispersion of $\sim$ 7.6 km s$^{-1}$.
These are the 'standard' HI clouds and have been well-studied for decades. 
However, we also
detect a second population of absorbing clouds with a velocity dispersion
of $\sim$ 21 km s$^{-1}$. About 20$\%$ of the total population of absorbing
clouds belong to this population. This new population of 'fast' clouds can 
be identified with a similar velocity dispersion Ca II absorbing clouds
and with the Halo clouds recently detected in HI emission from Green Bank.

\end{abstract}


\section{Introduction}

Observations carried out more than five decades ago 
of interstellar absorption lines of neutral sodium (NaI) and singly
ionized calcium (CaII) toward early type stars revealed several discrete absorption
features (Adams 1949, Blaauw 1952, Siluk and Silk 1974). 
These observations and their subsequent analysis led to the 
'cloudy' picture of the interstellar medium in which each of the discrete absorption
lines was associated with a concentration of gas. Subsequent HI 21 cm-line 
surveys of the Galaxy and theoretical studies led to the two-component model of the
interstellar medium in which dense, cold HI concentrations are in pressure
equilibrium with the diffuse, warm intercloud medium (Field, Goldsmith, \& Habing 1969, 
Clark, Radhakrishnan, \& Wilson 1962, 
Clark 1965, Radhakrishnan et al. 1972). The current picture
of the interstellar medium is the three component model where the third 
component is the tenuous, hot coronal gas created largely due to the supernova
activity (McKee \& Ostriker 1977, Wolfire et al. 1995).

Although a consistent picture of the ISM has evolved over the years, there are
some unresolved issues concerning the 'clouds' detected in the absorption lines
of NaI and CaII. Consider Fig. 1 which shows a histogram of the radial velocities
of discrete components observed in the direction of 64 stars from the data in Siluk and Silk (1974).
Since most of these stars are nearby ($< \sim$ 500 pc) the radial 
velocities of the absorbing features are essentially due 
to the random motions of the clouds.
The distribution of the low velocity features can be 
approximated well by the Gaussian shown
(Fig. 1). These
'slow' clouds were detected in HI 21 cm-line absorption and emission and have mean
values of HI column densities of $\sim$ 3 $\times$ 10$^{20}$ cm$^{-2}$, 
spin temperatures of $\sim$
80 K and volume densities of $\sim$ 20 cm$^{-3}$ (Spitzer 1978). 
There are $\sim$ 4 such concentrations
per kpc. These slow clouds have an effective scale height of $\sim$ 250 pc and 
are the 'standard' HI clouds. The higher velocity features (fast clouds) were not
detected either in HI absorption or in HI emission. These fast clouds also showed
anomalous ratios of the column density of sodium to singly ionized calcium. This ratio
was less than 1 for the fast clouds while significantly greater than 1 for the
slow clouds (Routly \& Spitzer 1952). 
If the fast clouds were shocked clouds, calcium which is normally locked
in grains will be released to the gas in such clouds, resulting in the observed
lower values of the NaI / CaII ratios. Such shocked clouds could be expected to
be of lower HI column density and of higher spin temperature resulting in lower
optical depth for absorption in the HI 21 cm-line ( Radhakrishnan \& Srinivasan
1980, Rajagopal, Srinivasan \& 
Dwarakanath 1998). Such fast clouds might also
go undetected in HI emission due to their lower column densities.

The motivation for the survey carried out with the Giant Meterwave Radio Telescope
(GMRT) was to detect such a population of fast clouds in HI 21 cm-line
absorption. The basic idea was to build a histogram for the local HI clouds similar to 
the one shown in Fig. 1. Such a study can compare the fast clouds detected so far
only in the absorption lines of NaI and CaII with those detected in HI 21 cm-line
absorption.

\section{Observations}

Extragalactic radio sources unresolved in the Very Large Array B-configuration 
and with their 21 cm flux densities greater than 1 Jy were selected. The magnitude
of the Galactic
latitudes of the sources were restricted to be more than 15$^{o}$ 
to avoid the confusion
caused by HI absorption in the plane and to enhance the chances of detection of local 
clouds in HI absorption. A total of 104 sources were selected which are distributed
rather uniformly in the Galactic longitude and latitude.
The observations were carried out using the GMRT
during the months of March-April, 2000 and April-June, 2001. About
120 hours of observations were carried out. At a velocity resolution of $\sim$
3.3 km s$^{-1}$ the detection limit in optical depth was $\sim$ 0.01 (3$\sigma$). 
This is the most sensitive high latitude HI absorption survey to-date.

\section{Results}

An example HI
absorption spectrum along with the corresponding HI emission spectrum from
the Leiden Dwingeloo Sky Survey (Hartmann \& Burton 1995) is shown 
in Fig. 2. Multiple Gaussian components were fitted to both the absorption and 
the corresponding emission spectra and discrete components were identified. 
All the spectra and the components identified are presented elsewhere. 
A total of 120 discrete absorption features were detected in the survey. A 
histogram of these features is compared with some of the earlier HI absorption
surveys in Fig. 3. The significantly larger number of HI absorption 
components detected in the present survey due to its lower detection limit
in optical depth is evident. 
It is also evident that the current survey has detected larger  number
of clouds at larger random velocities than the earlier surveys. 
A normalised histogram of the HI absorbing clouds from
the GMRT survey is shown in comparison to that of the data from Siluk and Silk (1974)
in Fig. 4. The histogram of HI clouds now clearly has a high velocity tail extending
up to $\sim$ 40 km s$^{-1}$ and looking more like the histogram of clouds
detected in optical absorption lines. 

The histogram of HI clouds is best fit by two Gaussians with velocity dispersions
of 7.6 $\pm$ 0.3 km s$^{-1}$ and 21 $\pm$ 4 km s$^{-1}$ respectively (Fig. 5).
The total number of fast clouds (V$_{lsr} >$ 15 km s$^{-1}$) is about 20\% of
the total number detected in the survey. The mean optical depth of these fast 
clouds is 0.04 $\pm$ 0.02 with a mean spin temperature of 125 $\pm$ 82 K and 
a mean HI column density of (4.3 $\pm$ 3.4) $\times$ 10$^{19}$ cm$^{-2}$.
The HI column density estimates were made from the spectra obtained from the 
Leiden Dwingeloo Sky Survey (Hartmann \& Burton 1995).
The column densities of these discrete components (N$_{HI}$) are plotted against
their velocities in Fig. 6. 

\section{Discussion}

The nature of the clouds in the high velocity tail first detected in the histogram 
of optical absorption lines (Fig. 1) is now clearer. The sensitive, high latitude,
HI 21 cm-line absorption survey from the GMRT has detected some of
these fast clouds. These fast clouds have three times 
larger velocity dispersion and ten times lower column
densities compared to the slow clouds as might be expected if they 
were from a shocked population of clouds.
The decrease in the HI column densities of the fast clouds as a function of their
random velocities (Fig. 6) is also consistent with this scenario. 
The shocked HI clouds are also
expected to be warmer than the slow clouds. The mean spin temperature of the fast 
clouds detected in the GMRT survey is not significantly higher than that 
of the standard slow clouds. However,
this might be a selection effect since for a given optical depth detection limit
and an HI column density, clouds with lower spin temperature will be preferentially
detected.

The fast clouds with three times higher dispersion are expected to have a scale height
about ten times larger compared to the slow HI clouds.
Given an effective thickness of 250 pc for the slow clouds, the fast
clouds can have an effective thickness of $\sim$ 2.5 kpc. The fast clouds can be 
part of the halo of the Galaxy. Recent HI emission studies using the Greenbank
telescope have led to the discovery of a population of discrete HI clouds 
in the Galactic halo with a velocity dispersion similar to that of the fast clouds
reported here (Lockman 2002). The fast clouds, once detected only in optical absorption lines,
have now been detected in both HI absorption and emission leading to a clearer
picture of the interstellar medium. Further details of the GMRT HI absorption survey
and the data analysis as well as a more elaborate discussion of the results are
found in Mohan, Dwarakanath \& Srinivasan (2004a,b).

\section*{Acknowledgements}
This article is a written version of the invited talk presented in 
the two-day
workshop held at the National Centre for Radio Astrophysics (Pune) during March 22 - 23, 2004 in honor of Prof. Govind Swarup's 
seventy fifth birthday. The results presented here form part of the PhD Thesis
of Rekhesh Mohan (2003).

\clearpage

\begin{figure*}
\epsfig{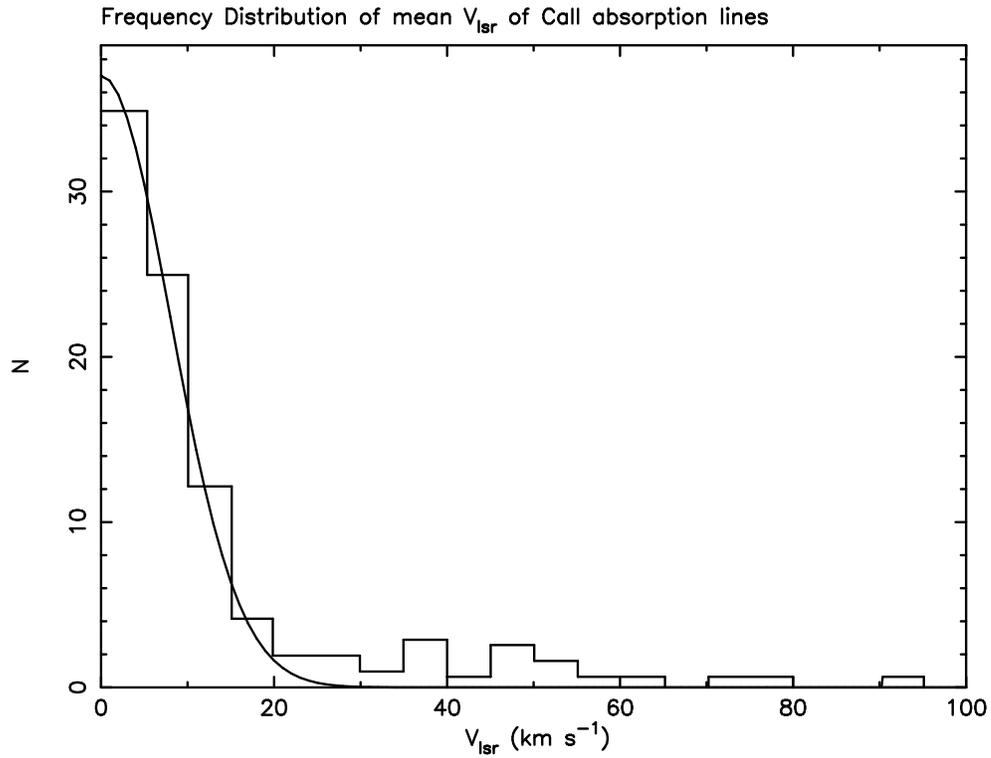}
\caption {Histogram of radial velocities observed in the direction of 64 stars
from the data in Siluk and Silk (1974). The low velocity features can be 
approximated well by the Gaussian shown ($\sigma$ = 8 km s$^{-1}$). 
The higher velocity features are associated
with anomalous ratios of sodium I to calcium II column densities. This figure is
reproduced from Radhakrishnan and Srinivasan (1980). The values of V$_{lsr}$ are
essentially the random velocities of the components since the stars are in the
solar neighbourhood.
 }
\end{figure*}

\begin{figure*}
\epsfig{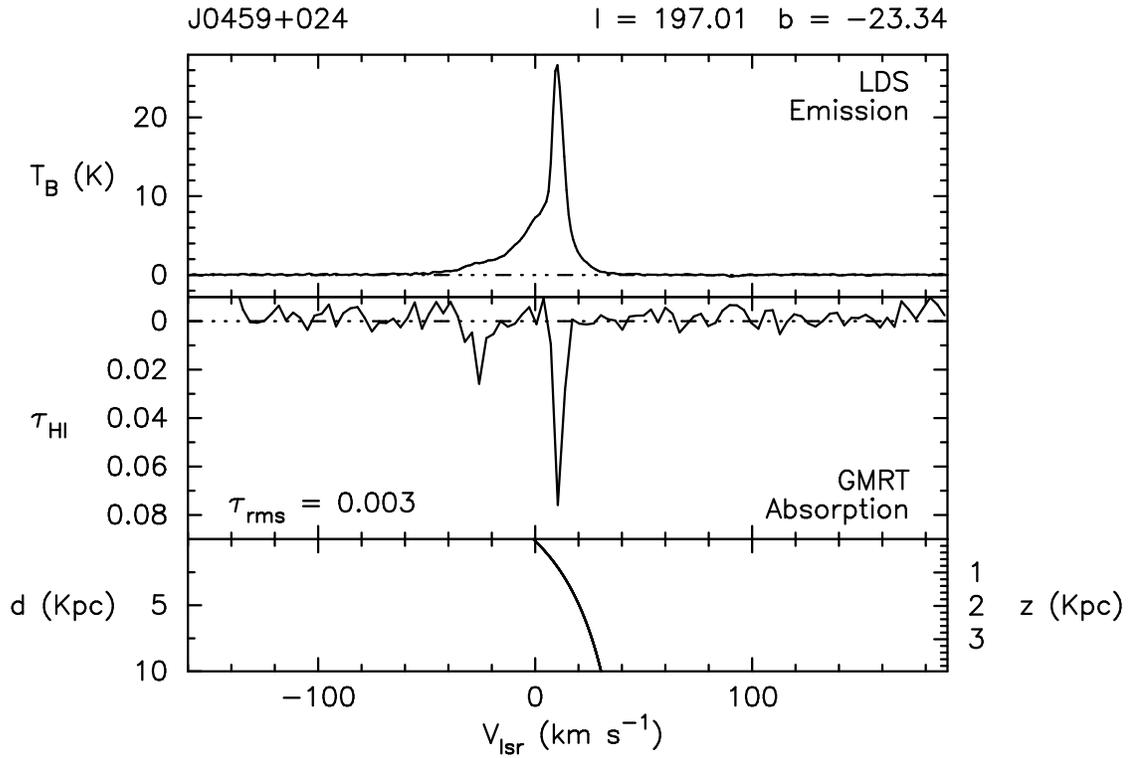}
\caption { A typical HI 21 cm-line absorption spectrum from GMRT (middle panel).
The top panel is the corresponding HI emission from the Leiden Dwingeloo Sky 
Survey (Hartmann \& Burton 1995). The bottom panel is the Galactic rotation curve 
for the given line of sight.
The heliocentric distance and the height above the plane are marked on the left and
the right sides of the bottom panel respectively. The absorption feature at V$_{lsr}
\sim$ --25 km s$^{-1}$ is an example of high random velocity feature. }
\end{figure*}

\begin{figure*}
\epsfig{file=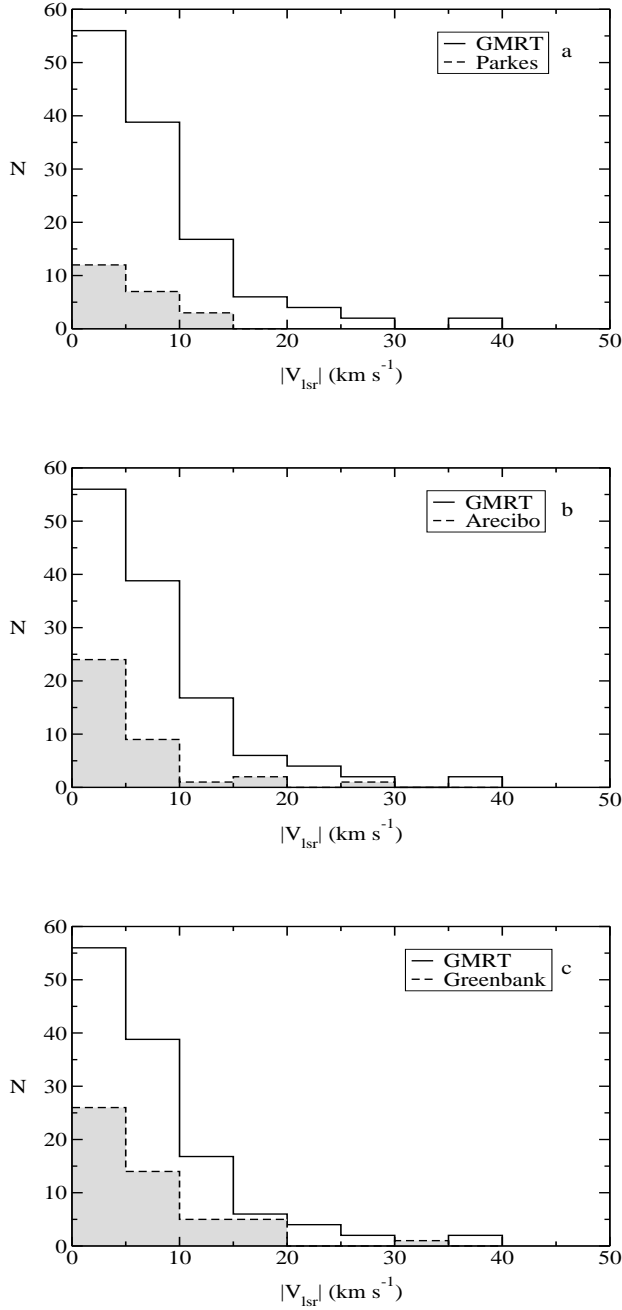, width=10cm}
\caption { Histograms of the number of HI absorption components from the GMRT survey
(solid line) compared
to those from the Parkes (Radhakrishnan et al 1972), the Arecibo (Dickey et al 1978)
 and the Greenbank (Mebold et al 1982) surveys. The larger number of components
detected in the more sensitive GMRT survey are evident. Furthermore, the GMRT survey
has detected more components with larger random velocities than the earlier surveys. }
\end{figure*}

\begin{figure*}
\epsfig{file=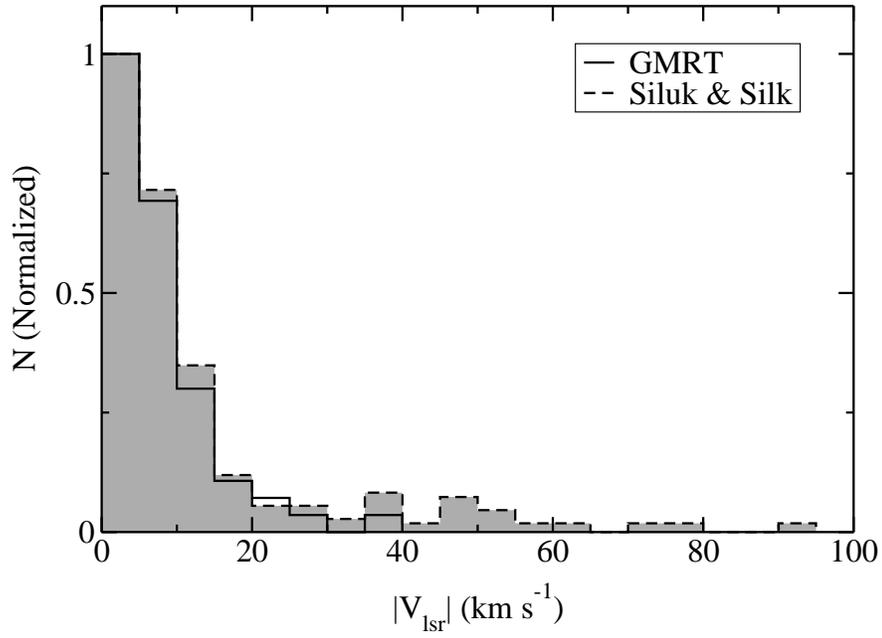, width=10cm, angle= -90}
\caption { Normalised histograms of the number of components detected in the GMRT 
survey (solid line) and those detected in the optical data by Siluk and Silk (1974). 
There is significant overlap
of the two histograms in the higher ($>$ 15 km s$^{-1}$) velocity region. 
}
\end{figure*}

\begin{figure*}
\epsfig{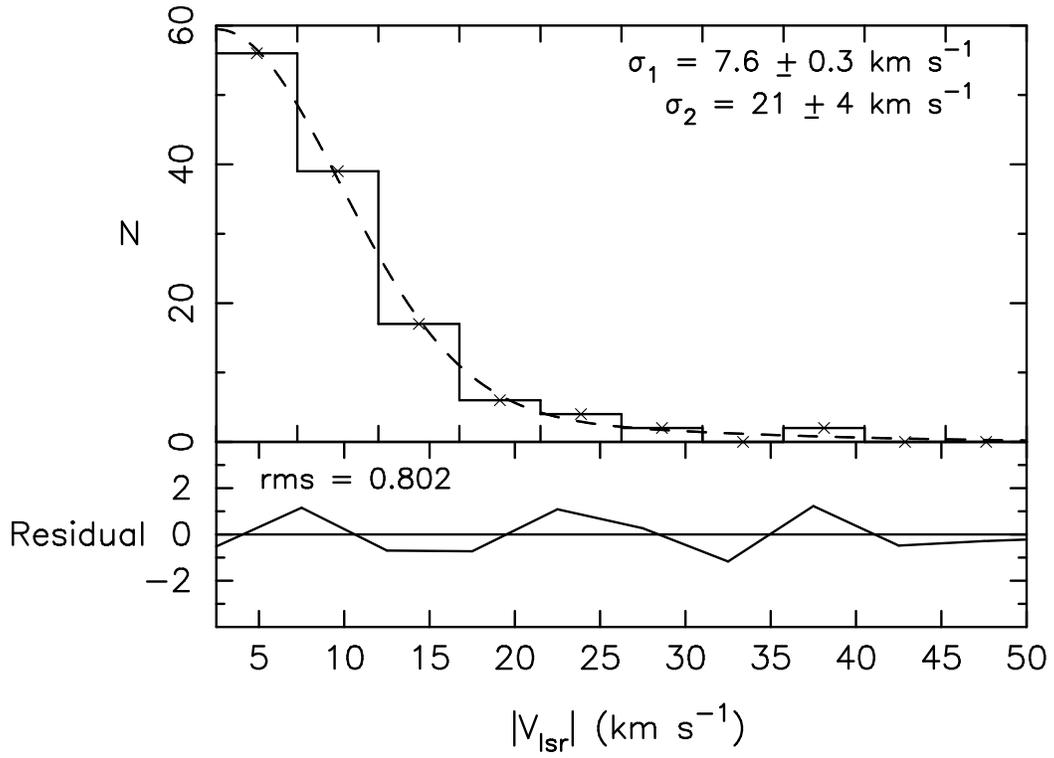}
\caption { Histogram of the discrete HI absorption components from the GMRT survey.
The broken line is a two Gaussian best-fit to the histogram. The bottom panel is the
residual. }
\end{figure*}

\begin{figure*}
\epsfig{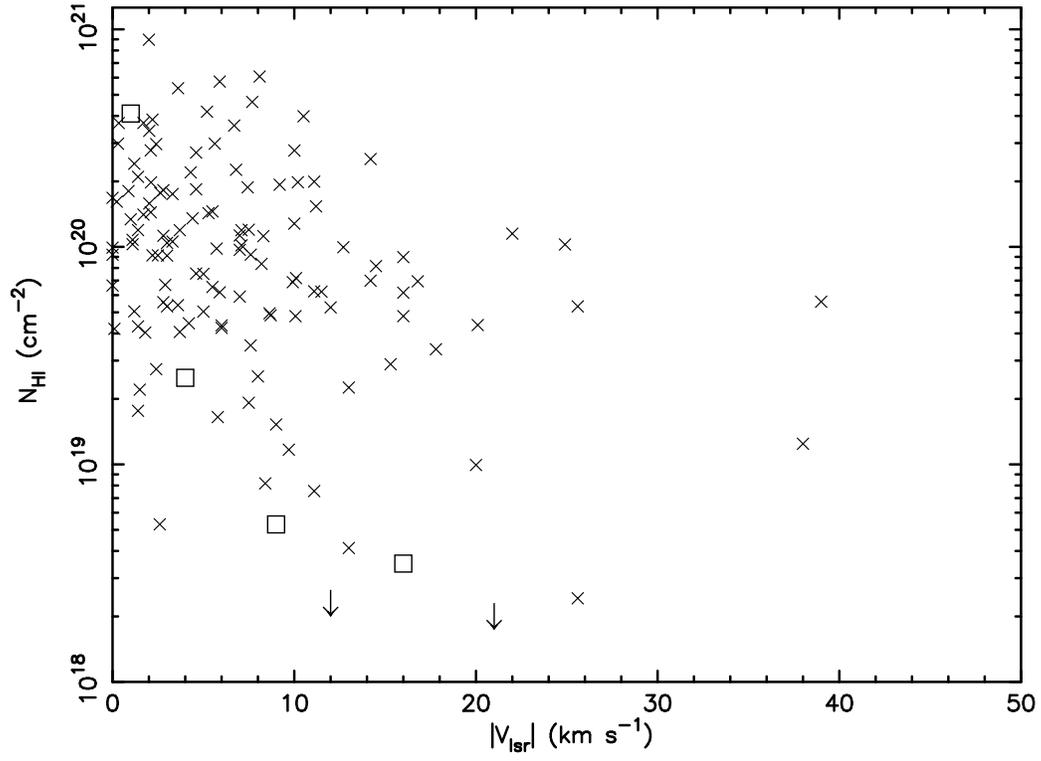}
\caption { Column density as a function of $\mid$ V$_{lsr}$ $\mid$. The mean HI column density
decreases with increasing random velocity. The squares are HI column density estimates
and the arrows are upper limits from the UV absorption line data (Martin and York 1982). }
\end{figure*}

\end{document}